# Terahertz emission from interdigitated photoconductive antennas based on Ge-on-Si

Dhanashree Chemate[1,2,3,*] Abhishek Singh[4,*], Ruturaj Puranik[2], Utkarsh Pandey[2], Dipti Gupta[1], Siddhartha P. Duttagupta[5], and Shriganesh S. Prabhu[2]

[1]Department of Metallurgical Engineering and Materials Science, Indian Institute of Technology, Bombay, Mumbai, MH, India, 400076

[2]Department of Condensed Matter Physics and Materials Science, Tata Institute of Fundamental Research, Mumbai, MH, India, 400005

[3]Centre for Interdisciplinary Sciences, Tata Institute of Fundamental Research, Hyderabad, TG, India, 500046

[4]Centre for Advanced Electronics, Indian Institute of Technology, Indore, MP, 453552

[5]Department of Electrical Engineering, Indian Institute of Technology, Bombay, Mumbai, MH, India, 400076

E-mail: dhanashree.c@iitb.ac.in, asingh@iiti.ac.in

**Abstract**

An interdigitated photoconductive antenna (i-PCA) for terahertz (THz) emission with a novel metal-insulator-semiconductor interface is designed with the aim of developing compact and scalable THz devices. The photoconductive material is an amorphous germanium (Ge) film deposited using DC magnetron sputtering. The antenna electrodes are composed of gold-germanium (AuGe). With the integration of a silicon dioxide ($SiO_2$) layer that acts as an electrical mask on alternate active areas, we present a simple approach to fabricate a large-area i-PCA. Along with a simplified fabrication compared to other existing designs, our approach increases the electrical robustness of the emitter and reduces the inactive gap area on the device. The i-PCA is capable of THz emission up to 2.5 THz and 36 dB signal-to-noise ratio (SNR), and is promising for applications in CMOS technologies.

## 1. Introduction

Photoconductive antennas (PCAs) are an integral part of the technological advancement of terahertz (THz) technologies. A typical PCA is composed of a semiconductor material with carrier lifetimes ranging from sub-picoseconds to sub-nanoseconds and a metallic antenna on top with various designs, from a single dipole to arrays of electrodes [1,2]. PCAs can be used both as a source and detector to facilitate coherent THz radiation and detection [3–5]. Its compact architecture, room-temperature operation, tunability over the THz output power, and polarization via applied bias and electrode design have led to its widespread adoption over the years [6]. However, with the growing demands in fields such as materials science, biomedical research, security applications, and high-speed communication, the development of scalable, cost-effective, and Si-CMOS-compatible THz sources and detectors remains crucial [7,8].

In source PCA, when the inter-electrode gap is illuminated by a femtosecond (fs) optical pulse with an energy above the semiconductor bandgap, it generates electron-hole pairs that are accelerated by the applied voltage. Consequently, a transient current is produced, and the output radiation has an electric field proportional to the time derivative of the current, that is, $E_{THz} \propto \frac{\partial J}{\partial t}$ [9]. The important performance parameters for PCAs are the THz power, spectral bandwidth, signal-to-noise ratio (SNR), and both electrical-to-THz conversion efficiency and optical-to-THz conversion efficiency. Although the choice of photoconductive (PC) material primarily governs the above characteristics [2], the electrode architecture also plays a critical role [10,11].

Traditional designs for THz PCAs include single dipoles, such as bowtie, Hertzian, and Stripline, which are easy to fabricate and avoid phase interference owing to their single-point source configuration [1,12]. However, dipole antennas suffer from two key challenges: saturation caused by the Coulomb screening effect and thermal breakdown due to optically induced heating [13], which essentially limits their operation to lower optical excitation powers and voltage biases. These limitations have motivated the exploration of alternative geometries. One such design is the interdigitated electrode, known as interdigitated PCA (i-PCA). It typically consists of a parallel arrangement of multiple anodes and cathodes intertwined with each other [14]. Owing to its large-area architecture, the local heating and saturation effects are minimized [15]. Additionally, the alignment of the pump beam is easier than that of the single dipoles, as the entire device area, with a range of a few micrometers to millimeters, is illuminated. In brief, this structure offers a relatively simple approach, both in fabrication and practicality.

However, one limitation of i-PCA is that the bias electric field in the adjacent gaps has the opposite direction, which leads to net destructive interference in the output electromagnetic (EM) field. To address this issue, researchers have demonstrated shadow masking of alternate electrodes with metal, thereby making the EM field unidirectional [16]. Alternatively, as demonstrated in [17], trenches can be created by etching the photoconductive material from alternate gaps. However, the filling factor in the shadow mask is relatively low, and with the etching, the fabrication process leans towards complexity. Microlens arrays have also been used to selectively photoexcite alternate electrode gap areas [17,18]. Cylindrical and spherical microlens arrays have been used in several commercial THz emitters. However, the alignment of microlens arrays with i-PCA electrodes is extremely critical. Selective incorporation of plasmonic structures and nanostructure grating lines to i-PCA electrodes has also been used to achieve enhanced THz emission only from alternate active regions in i-PCAs [19–21]; however, these structures often require hybrid lithographic processes, such as E-beam combined with UV lithography, which limits the scalability.

To address the above issues, instead of a regular metal-semiconductor interface as shown in Fig. 1(a), we introduce a novel design for i-PCA. As illustrated in Fig. 1(b) and Fig. 1(c), the active and inactive regions of the device have different architectures. We incorporated silicon dioxide ($SiO_2$) as an electrical mask in the alternate gaps in i-PCA to suppress the THz emission and avoid destructive THz interference in the far field. Although photoexcitation can still occur in the inactive gap of the semiconductor beneath the $SiO_2$ layer, photogenerated charge carriers cannot be efficiently collected by the electrodes. This obstruction is expected to drastically affect the THz emission from inactive gaps. In our work, gold germanium (AuGe) is used as the electrodes, and sputtered Germanium (Ge) thin films as the PC material. Sputtering is a cost-effective and CMOS-compatible alternative to the complex epitaxial growth methods that are used for conventional PC materials. Furthermore, Ge offers several advantages that make it a suitable candidate for THz PCA applications: (i) the absence of polar phonons, which avoids spectral absorption dips and enables broader bandwidths [22,23] (ii) compatibility with a wide range of laser excitation (including 800 nm and fiber-lasers up to 1550 nm) and (iii) naturally short carrier lifetimes in Ge thin film (~ sub-ps) owing to its amorphous nature, having trapping sites that remove the need for ion implantation [24,25]. And because our design employs only one metallic layer, the second metallic layer, which is used to cover the alternate electrode gap region, is not required. Thus, a simplified fabrication process with one less lithography step is possible. The $SiO_2$ layer partially separates the electrodes and semiconductor materials, reducing the metal-semiconductor contact area and thereby decreasing the amount of dark current flowing through the device.

Overall, the combination of a novel i-PCA electrode design and an amorphous Ge material platform aligns with the broader goal of efficient THz sources for scalable CMOS-integrated THz systems. In the following sections, we discuss device fabrication, experimental setup, and the emission response from the device under study – a germanium-based interdigitated photoconductive antenna, referred to as Ge-i-PCA.

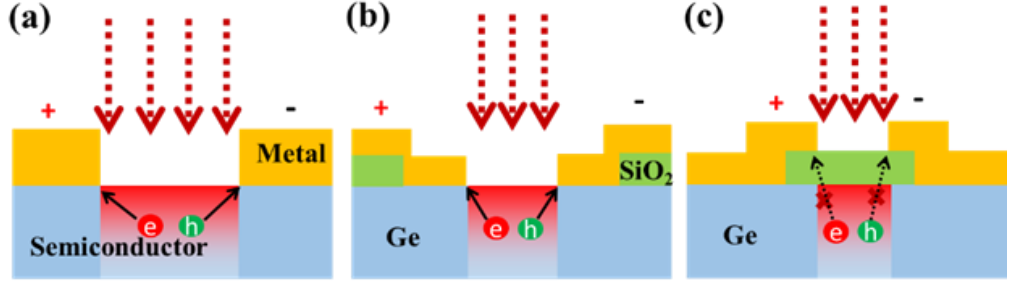

**Figure 1.** (a) Usual metal-semiconductor interface used for regular PCAs. (b) Cross-section of a strip line pair at the active gap in the proposed i-PCA design. The active gap area has a metal-semiconductor (M-S) interface at the inner edges of the strip line pair and an M-I-S interface at the outer edges of the strip line pair. Photogenerated charge carriers can easily reach the electrodes due to direct M-S contact. (c) Cross-section of a strip line pair with metal-insulator-semiconductor (M-I-S) interface at the inactive gap in the proposed i-PCA design. The insulating layer electrically masks the semiconductor from the electrode and blocks the direct path of the photogenerated charge carriers to the electrodes. Thus, suppressing the THz emission from alternate gaps (i.e., inactive gap) in i-PCA.

## 2. Device Fabrication and Simulations

The architecture of the device is illustrated in Fig. 2(a), with the detailed dimensions as follows: the active and inactive (electrically) area gaps between the electrodes are 10 μm and 5 μm, respectively; the width of the electrodes is 17.5 μm (strip line pair period is 50 μm), and the width of the $SiO_2$ bars beneath the electrodes is approximately 18 μm. In i-PCAs, the electrodes play a key role in acting as an antenna in THz emission. THz emission occurs from the semiconductor (i.e., the active gap) and from the electrodes, which act as an antenna. The emission is highly dependent on the electrode widths. However, the inactive gap area in an i-PCA does not contribute to THz emission. Hence, the inactive gap area (not the electrodes) is a region that should be minimized/optimized. In the case of a high bias voltage or low resistivity of the semiconductor, a narrow inactive gap area can lead to device breakdown. However, in our design, the use of a metal-insulator-semiconductor (M-I-S) interface at the inactive gaps makes the electrodes well insulated from each other and reduces the chances of electrical breakdown via the inactive gap. This novel design reduces inactive gaps to much narrower than active ones. In our device, we reduced the inactive gap to 50 % of the active gap, which is otherwise usually as wide as the active gaps. Since $SiO_2$ is electrically insulating with a really large dielectric breakdown of (>1 V/nm) [26], the inactive gaps can be further reduced depending on the fabrication limitation. We have utilized a research lab-grade UV photolithography system having a minimum feature size of ≤ 5 μm for multilayer lithography.

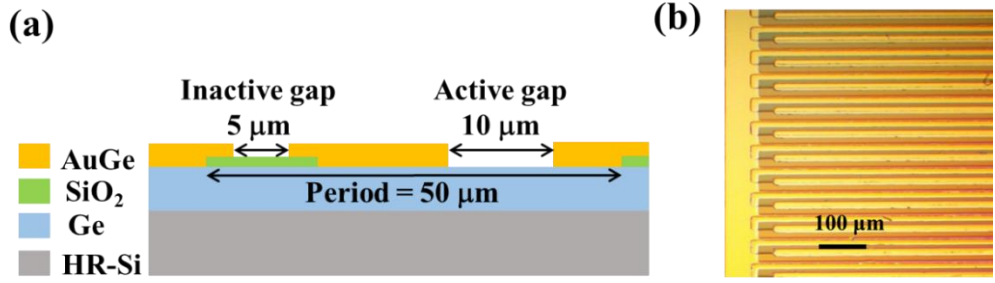

**Figure 2.** (a) The device architecture of Ge-i-PCA: The substrate used is a high-resistive silicon (HR-Si) with the Ge film as a photoconductive material on top and $SiO_2$ bars in the alternate electrodes of AuGe, with the corresponding dimensions as shown. (b) The optical microscopic image (magnified view) of the Ge-i-PCA; the $SiO_2$ layer is situated in the middle of the metallic electrodes in the alternate region. The total device area is 1 mm$^2$.

We used DC magnetron sputtering for the deposition of Ge on a high-resistive silicon substrate (resistivity of ~ 10000 Ω-cm). With a sputtering pressure of $8 \times 10^{-3}$ mbar and a deposition rate of ~ 0.44 nm/s, a 700 nm Ge film was deposited. Next, to fabricate the i-PCA, a negative photoresist was used to define the $SiO_2$ bar structures on top of the Ge film, and patterning was carried out using an iron oxide mask. We employed plasma-enhanced chemical vapor deposition (PECVD) to deposit a 160 nm-thick $SiO_2$ layer. A lift-off process was carried out to create periodic $SiO_2$ bars over the Ge layer. Subsequently, an additional layer of photoresist was patterned by UV exposure using a different mask. Following the development of the resist, AuGe with a thickness of approximately 270 nm was sputtered, and a lift-off process was used to define the interdigitated electrodes. Fig. 2(b) shows an optical microscopic image of a part of the fabricated i-PCA on a Ge thin film. In the end, to mitigate the risk of short circuits induced by dust or other contaminants, a protective layer of $SiO_2$ (160 nm) was deposited using an additional photolithography process employing a square-shaped iron-oxide mask. Subsequently, the final lift-off process completed the formation of a multilayer structure.

The role of $SiO_2$ in our i-PCA design can be understood through electric field simulations. Fig. 3(a) illustrates the simulated electric field for the case where there is no $SiO_2$ layer (i.e., regular M-S interface), and Fig. 3(b) shows the results for the M-I-S interface with a $SiO_2$ layer. The electric field is observed on a plane parallel to the device surface at a depth of 400 nm inside Ge, with an applied bias of 20 V. A significant drop in the field, especially near the electrode edge region, is seen when the $SiO_2$ layer is present. This is expected, as in the case of $SiO_2$, most voltage drops would occur across the thickness of the $SiO_2$ itself. Thus, the reduction in the bias field together with an increase in the path length for the photogenerated carriers (due to the presence of the SiO$_2$ layer at the inactive gap) would significantly suppress the THz emission from this area.

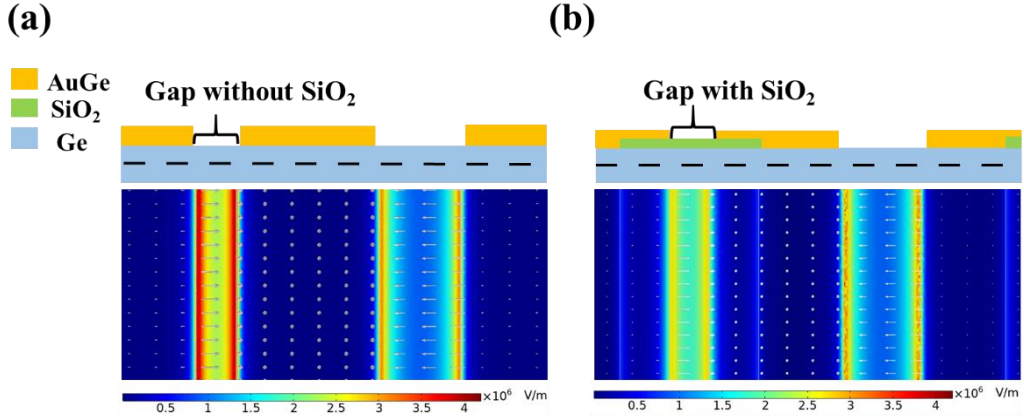

**Figure 3.** Simulated electric field observed inside Ge at 400 nm depth (shown as black dashed line) at a bias voltage of 20 V using COMSOL 5.6 Multiphysics software (a) Without SiO$_2$ layer (b) With the SiO$_2$ layer at the inactive gap. The presence of a SiO$_2$ layer between the electrodes and the Ge significantly reduces the bias field inside the Ge.

## 3. Experimental setup

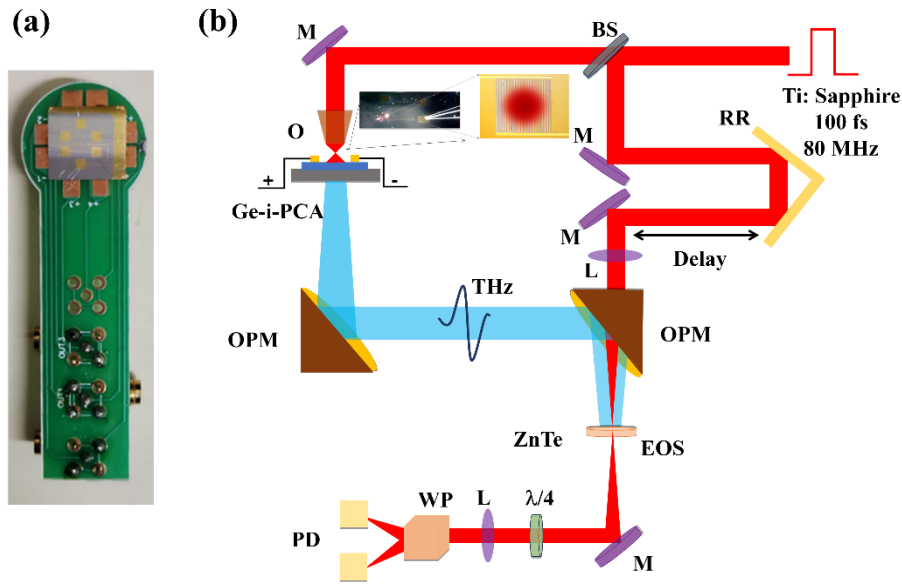

**Figure 4.** (a) The Ge-i-PCA device mounted on the PCB. (b) The THz time-domain setup utilized for the characterization of Ge-i-PCAs involves illumination with a 10x objective. The inset is an image of the illuminated device with a magnified depiction on the right. The device is biased at 20 V, and a chopping frequency of 27.437 kHz is utilized. Two parabolic mirrors are employed to collimate and focus the THz radiation generated from the i-PCA onto the ZnTe detector. A probe beam, with a power of 30 mW, is delayed for the subsequent EO sampling. M: Mirror, BS: Beam Splitter, RR: Retroreflector, L: Lens, WP: Wollaston Prism, PD: Balanced Photodiode, OPM: Off-axis Parabolic Mirror; the signal from the balanced photodiode is further fed into the lock-in amplifier.

The device was integrated and wire-bonded on a custom-made printed circuit board (PCB), as shown in Fig. 4(a), which was mounted onto a kinematic mirror mount. Next, to characterize the device, we used a mode-locked femtosecond laser (Tsunami from Spectra-Physics, with a central wavelength of 800 nm, a repetition rate of 80 MHz, and a pulse width of ~ 100 fs). The experimental setup is shown in Fig. 4(b). A 1 mm-thick <110> ZnTe layer was used for electro-optic (EO) sampling at the detector position. The device was mounted such that the electrodes were perpendicular to the pump beam. The pump beam was sent through a 10x objective, and the device was positioned away from the focus to ensure that the 1 mm² device area was uniformly illuminated.

## 4. Results and Discussion

The current–voltage (I–V) characteristics of Ge-i-PCA are shown in Fig. 5(a). The dark current in the device exhibits a linear dependence on the voltage. This linearity suggests ohmic behavior with high-quality metal-semiconductor contacts. The magnitude of the current through the device is approximately 1.2 mA at a bias voltage of 20 V, which is quite low despite having several pairs of stripline electrodes on Ge. Typically, devices fabricated on Ge suffer from large dark currents and subsequent heating. A single bowtie antenna with a 10 µm electrode gap on bulk Ge has been reported to produce a dark current of ~ 7 mA at a bias voltage of 10 V [27]. Thus, our sputtered Ge film provides an electrically robust alternative to bulk Ge, enabling the fabrication of larger interdigitated electrode-based photoconductive emitters with a reasonably low dark current. Our M-I-S interface-based i-PCA electrode design should also have contributed to reducing the dark current. A lower dark and photocurrent implies that the electrical energy consumed by the device is less, and it could result in a better electrical-to-THz conversion efficiency. Although providing higher electrical energy is comparatively much easier and more economical than providing higher optical pump energy to the THz emitter, managing the heat generated by the electrical energy consumption in the emitter is challenging and usually limits the overall THz emission performance of the emitter.

Fig. 5(b) illustrates the photocurrent (i.e., the total current under illumination minus the dark current) characteristics of the device under optical excitation powers of 30 mW (and 35 mW to show repeatability under similar optical excitations). The photocurrent is ~ 1 mA at 20 V and 0.5 mA at 10 V bias at a pump power of 35 mW. Thus, the photocurrent through our sputtered Ge device is also much lower (by an order of magnitude) than that of the bulk Ge-based devices [27]. The lower photocurrent can be attributed to the shorter carrier lifetime and lower mobility of the amorphous Ge, which is typically $10^{-3} – 10^{-2}$ cm²/V-s [28]. A lower mobility is not suitable for efficient THz emission; however, a shorter carrier lifetime in sputtered Ge is not detrimental to THz emission. Instead, it can increase the THz emission as well as the detection efficiency [29].

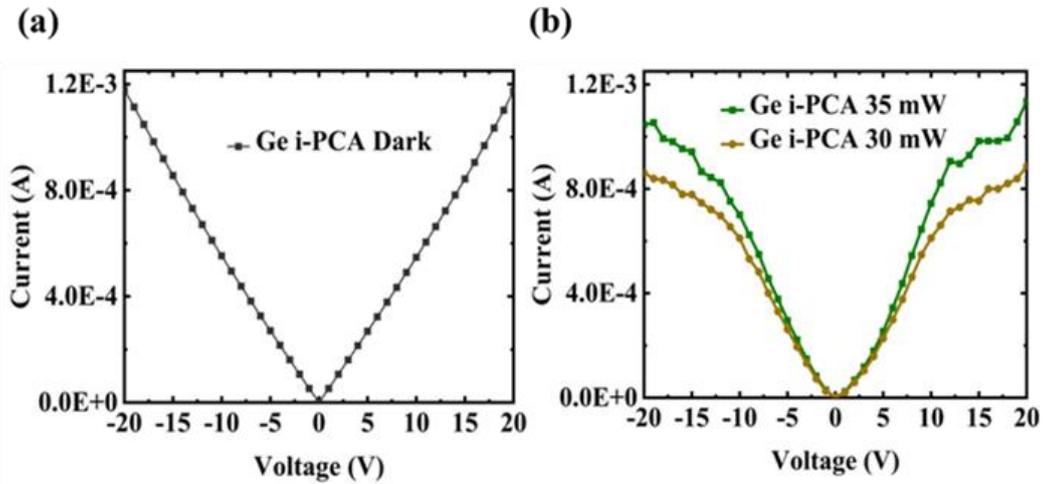

**Figure 5.** (a) I-V characterization of the Ge-i-PCA (a) Dark current and (b) Photocurrent at 30 mW and 35 mW

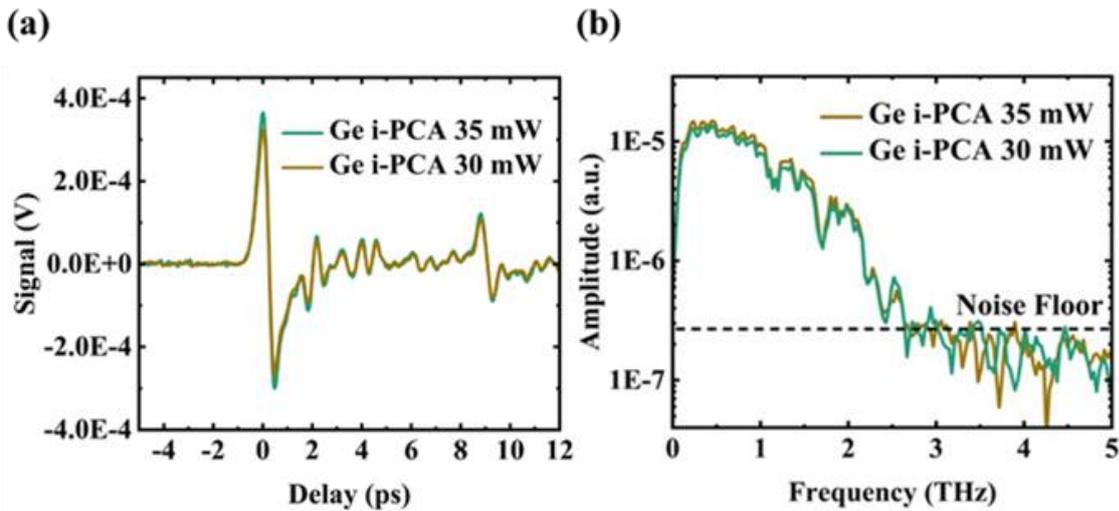

**Figure 6.** (a) THz time domain response of Ge-i-PCA at an illumination of 30 and 35 mW with a bias of 20 V. (b) Fourier Transform of the THz time domain pulse with the bandwidth reaching up to 2.5 THz and an SNR of 36 dB

Next, the devices were tested for THz emission using a time-domain spectroscopy setup, as shown in Fig. 4(b). The recorded time-domain THz pulse emitted from the device is shown in Fig. 6(a). Under both (30 mW and 35 mW) optical conditions, we observed a similar THz pulse, with the 35 mW pump condition having a slightly higher THz field, as expected. The corresponding frequency-domain analysis, shown in Fig. 6(b), presents the Fourier transform of the THz time-domain signal. The spectral bandwidth of this device extends up to 2.5 THz with an SNR of 36 dB, exhibiting THz generation from an i-PCA on Ge thin film with a novel

AuGe-SiO$_2$-Ge (M-I-S) interface. While ion-implanted crystalline Ge holds the record for the THz bandwidth, this approach requires complex and costly post-growth processing. We demonstrate that a simple, low-cost, sputtered amorphous film can achieve a bandwidth of 2.5 THz. The SNR in our device is smaller compared to the state-of-the-art PCAs based on bulk Ge (54 dB [23]) and InGaAs (137 dB [30]), mainly because of its low carrier mobility.

However, to the best of our knowledge, this represents the highest bandwidth reported for this specific class of material, i.e., germanium thin film [22,31], positioning it as a practical and scalable alternative for applications where the ultimate bandwidth is not the primary concern. THz emission can be significantly enhanced by improving the Ge film quality and hence the carrier mobility. We note that the measurements were performed without the integration of a hemispherical silicon lens, which is commonly employed to mitigate total internal reflection losses at the semiconductor-air interface; hence, there is further scope to improve the SNR and, consequently, the bandwidth. Additionally, the experiments were conducted in ambient air. A nitrogen-purged environment would be more suitable for such sensitive measurements and is expected to enhance the SNR further. Furthermore, the Joule heating in our device is expected to be quite low compared to that in the bulk Ge device, as the total current flowing through the device remains limited to approximately 2 mA at a bias voltage of 20 V. Thus, there is a possibility of further improvement in the THz emission by increasing the bias voltage of the emitter. Another key aspect during the characterization of THz sources is the pulse width of the pump laser. A shorter pulse-width pump laser, instead of a 100-fs laser, can be utilized for a faster rise of photo-induced charge in order to improve the photocurrent, thus possibly leading to a broader bandwidth [23].

Finally, the stability of device performance was evaluated by performing a bidirectional voltage sweep, as shown in Fig. 7. The device gives a stable output with little to no change in the THz peak values when the voltage is swept forward and in the reverse direction, reflecting the reliability of the device for long-term and repetitive operation. The THz field dependence on the applied bias appears slightly superlinear between 10 V and 14 V bias instead of linear, which is not ideal for a photoconductive THz emitter. As discussed in [27], superlinearity is not an inherent characteristic of Ge-based emitters but is likely influenced by the nature of the contacts. In our case, we speculate that our M-I-S interface may have contributed to this. Notably, at biases lower than 10 V and higher than 14 V, it is almost linear, as it should be.

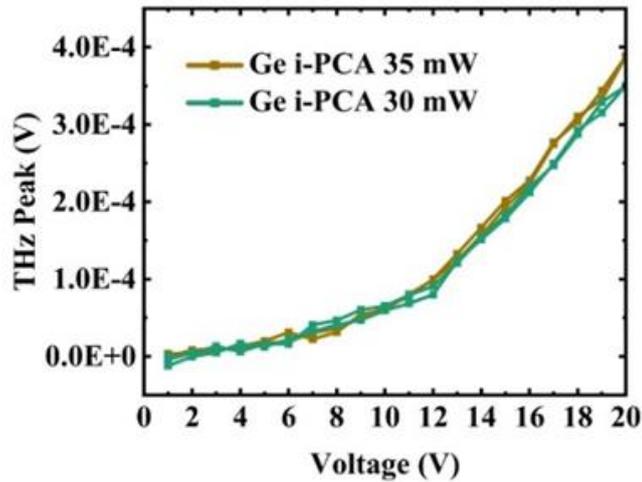

**Figure 7.** THz peak vs Voltage with a bidirectional sweep showing the repeatability of the device

## 5. Conclusion

We have presented a novel design for an i-PCA utilizing an M-I-S interface to enhance electrical robustness and simplify fabrication for THz emitters. We demonstrated the THz emission from amorphous Ge-on-Si substrates. By incorporating $SiO_2$ into the alternate electrode, we could reduce the inactive gaps to 50% of the active gaps in our i-PCA. With the sputtering of Ge, a simple and alternative approach for PC material growth is explored. The device could reach up to 2.5 THz in bandwidth with an SNR of 36 dB, with a further scope of improvement by using a silicon lens and carrier mobility by improving the Ge film quality. This design is suitable for compact, cost-effective, and scalable THz sources that can be integrated into emerging Si-CMOS technologies.

**Acknowledgements**

The authors thank Dr. Gajendra Mulay and Mr. Amit Shah for their help with $SiO_2$ deposition.

**Funding**

Department of Atomic Energy, India, vide grant RTI4003 and SERB India under project RJF/2022/000081.

**Disclosures**

The authors declare no conflicts of interest.

**Data availability**

Data underlying the results presented in this paper are not publicly available at this time but may be obtained from the authors upon reasonable request.